\documentclass[11pt]{article}

\usepackage[margin=1in]{geometry}
\usepackage[T1]{fontenc}
\usepackage[utf8]{inputenc}
\usepackage{newtxtext,newtxmath}
\usepackage{microtype}
\usepackage{amsmath}
\usepackage{booktabs, tabularx, array}
\usepackage{enumitem}
\usepackage{xcolor}
\usepackage{xurl}
\usepackage{abstract}
\usepackage{titling}
\usepackage{titlesec}
\usepackage{fancyhdr}
\usepackage{graphicx}
\usepackage{tikz}
\usepackage{hyperref}
\usepackage[numbers,sort&compress]{natbib}
\usetikzlibrary{arrows.meta, positioning, fit, shapes.geometric, calc}

\definecolor{paperblue}{HTML}{17324D}
\definecolor{papergray}{HTML}{5E6B76}
\definecolor{providerred}{HTML}{A63D40}
\definecolor{deployergreen}{HTML}{2E6E56}
\definecolor{softblue}{HTML}{EAF1F7}
\definecolor{softred}{HTML}{F8EAEA}
\definecolor{softgreen}{HTML}{E8F3EE}

\hypersetup{
  colorlinks=true,
  linkcolor=paperblue,
  citecolor=paperblue,
  urlcolor=paperblue,
  pdftitle={Final Authority in AI Governance},
  pdfauthor={Zexun Wang},
}

\setlist[itemize]{leftmargin=1.4em}
\setlist[enumerate]{leftmargin=1.6em}
\titleformat{\section}
  {\Large\bfseries\color{paperblue}}
  {\thesection.}
  {0.6em}
  {}

\titleformat{\subsection}
  {\large\bfseries\color{paperblue}}
  {\thesubsection}
  {0.6em}
  {}

\titlespacing*{\section}{0pt}{1.4em plus 0.2em minus 0.2em}{0.6em}
\titlespacing*{\subsection}{0pt}{1.0em plus 0.2em minus 0.2em}{0.35em}

\pretitle{\begin{center}\LARGE\bfseries\color{paperblue}}
\posttitle{\par\end{center}\vspace{0.35em}}
\preauthor{\begin{center}\large}
\postauthor{\par\end{center}\vspace{-0.3em}}
\predate{\begin{center}\small\color{papergray}}
\postdate{\par\end{center}\vspace{0.8em}}

\newcommand{\paperkeywords}[1]{%
  \begin{center}
    \small\textbf{Keywords:} #1
  \end{center}
  \vspace{0.4em}
}

\title{
  Final Authority in AI Governance:\\
  Frontier-Provider Sovereignty and\\
  Action-Centered Deployer Governance
}
\author{Zexun Wang\\Ond Holdings Inc.\\Toronto, Canada\\\texttt{jw@nd.im}}
\date{June 2026}

\begin{document}

\maketitle

\begin{abstract}
This paper examines where final authority should sit once capable AI systems are embedded in real organizational workflows. It compares two governance models. The first, \emph{frontier-provider sovereignty}, assigns privileged authority to the provider of the most capable models and is reflected in contemporary arguments for frontier-model testing, release gating, transparency duties, and compute-related controls \citep{amodei2026policy, amodei2025deepseek}. The second, \emph{action-centered deployer sovereignty}, places final authority over high-impact actions with the organization that authorizes the action, embeds it in a business process, and bears the downstream legal, operational, and commercial consequences.

Methodologically, the paper combines comparative reading of public governance frameworks with implementation-informed analysis of runtime heterogeneity and enterprise control requirements. The comparison covers EU AI Act guidance, the NIST AI Risk Management Framework, Singapore's Model AI Governance Framework for Agentic AI, recent Japanese AI policy instruments, and Canada's voluntary code and managerial guidance \citep{eu2025gpai, eu2026art14, eu2026art26, nist2023airmf, nist2024genai, imda2026agentic, japan2026guidelines, japan2025basicplan, japan2025act, canada2023voluntary, canada2025guide}. Across these materials, the paper finds stronger support for distributed operational accountability than for unilateral frontier-provider control. It further shows that rapid enterprise adoption, declining provider transparency, and widening control gaps increase the value of a portable governance layer centered on governed action rather than on provider-native session objects \citep{stanford2026index, ibm2026controlgap, mckinsey2026trust, cisco2025readiness}. The resulting conclusion is layered rather than absolutist: strong upstream authority remains justified for frontier capability gating, but final authority over concrete enterprise action is better located with the deployer and consequence-bearer. Under contemporary enterprise conditions, PCAA-like route-review-prove systems---derived from the prior \emph{Proof-Carrying Agent Actions} formulation centered on governed actions and replayable proof---appear better aligned with enterprise governance requirements than provider-centric theories of AI sovereignty \citep{wang2026pcaa}.
\end{abstract}

\paperkeywords{AI governance, agent governance, frontier models, enterprise AI, Proof-Carrying Agent Actions, runtime governance}

\section{Introduction}

Debates about AI governance are often presented as disputes over safety versus acceleration, or over open access versus responsible restraint. Those distinctions are politically salient, but they leave an underlying institutional question under-specified. Once AI systems become capable of taking or mediating consequential actions inside organizational workflows, where should final authority sit? This paper treats that question as one of \emph{governance sovereignty}: who should hold the decisive right to authorize, delay, escalate, deny, and later justify high-impact AI action.

This question matters because modern AI systems no longer remain confined to the model layer. They are now embedded in agent runtimes, managed infrastructures, tool ecosystems, enterprise data planes, and long-horizon workflows. Anthropic offers managed agent harnesses running in hosted infrastructure. Microsoft offers a managed platform for deploying and scaling agents. LangSmith offers trace-centric observability across many providers and frameworks. NVIDIA offers programmable guardrails as an application-layer control surface \citep{anthropic2026managed, microsoft2026foundry, langsmith2026observability, nvidia2026guardrails}. These are not signs of a stable, unified control stack. They are signs of a heterogeneous ecosystem in which the same governed action may appear as a model call, a runtime transition, a tool invocation, an approval event, a trace span, or a post hoc audit artifact.

In such a landscape, the location of final authority becomes more important, not less. A provider can legitimately say that it knows the model family best. A runtime can legitimately say that it sees execution most directly. An observability product can legitimately say that it preserves the richest trace. But none of those claims alone resolves the enterprise question of authority. Someone still has to decide whether a governed action may proceed, whether it requires review, whether a hold is pre-execution or merely advisory, and what evidence must remain after the action completes.

The paper evaluates two rival answers through a comparative governance analysis that combines three kinds of material: public framework documents, provider policy statements, and implementation-visible evidence about enterprise runtime heterogeneity. The objective is not to adjudicate all questions of AI safety in general. It is to identify which sovereignty model is better aligned with the governance requirements of enterprise AI action, and under what boundary conditions the rival model still retains legitimate force.

The first answer may be called \emph{frontier-provider sovereignty}. On this view, the provider of the most capable models should enjoy privileged governance standing because it best understands the model's capabilities, failure modes, and catastrophic-risk profile. In contemporary practice, this view is represented clearly by recent policy writing associated with Dario Amodei and by Anthropic's Responsible Scaling Policy. That body of argument holds that frontier AI models should be subjected to technical testing and auditing and that release should be blocked or reversed when sufficiently high safety standards are not met. It also supports stronger transparency requirements for frontier AI companies and stronger export controls on advanced chips \citep{amodei2026policy, amodei2025deepseek}. Anthropic's Responsible Scaling Policy, in turn, formalizes capability thresholds, risk reports, executive approvals, and in some cases external review for frontier-risk decisions \citep{anthropic2026rsp}. Read charitably, this is a coherent theory of upstream governance: the frontier provider should be trusted to know the danger first, to see the capability frontier earliest, and therefore to hold unusually strong gatekeeping power.

The second answer may be called \emph{action-centered deployer sovereignty}. On this view, provider safety remains valuable, but final authority over consequential AI action should sit with the organization that authorizes the action, embeds it in a real workflow, and bears the consequences when it goes wrong. In practical terms, that means the enterprise, agency, hospital, bank, law firm, or infrastructure operator that decides whether a governed action should be routed, reviewed, approved, denied, deferred, or recorded under explicit evidence semantics. One implementation-oriented form of this second theory is \emph{Proof-Carrying Agent Actions} (PCAA), introduced in prior work as a runtime-neutral governance model centered on an action certificate rather than on a vendor-native session record \citep{wang2026pcaa}. Its core move is to treat the governed action itself, rather than the provider session or runtime object, as the primary trust-bearing object.

The paper's thesis is narrower than a general case for ``decentralization'' and narrower than a general critique of frontier labs. It is that \emph{provider safety should constrain what models can do, but provider sovereignty should not monopolize who decides whether a governed enterprise action may occur}. In enterprise settings, the more defensible sovereign is ordinarily the consequence-bearer rather than the frontier model lab.

This paper makes five contributions. First, it separates \emph{provider safety} from \emph{provider sovereignty}, a distinction that is often blurred in public debate. Second, it introduces a comparative vocabulary for analyzing AI governance as a question of authority allocation rather than merely as a catalog of safety controls. Third, it compares major public frameworks in the EU, United States, Singapore, Japan, and Canada through that lens. Fourth, it argues that an action-centered governance primitive such as PCAA is better aligned with contemporary enterprise governance requirements than a provider-centric sovereignty model. Fifth, it clarifies the strongest surviving case for upstream authority and uses that clarification to defend a layered settlement rather than a simplistic anti-provider inversion.

\section{Analytical Lens and Method}

This paper is a qualitative comparative governance analysis rather than a causal identification study. Its unit of analysis is not the model alone, nor the firm alone, but the \emph{locus of final authority over consequential AI action}. The core question is institutional: when an AI-mediated action can change a production system, disclose information, bind an organization externally, or otherwise create material downstream consequence, where should legitimate final authority sit?

The analysis combines four source families. First, it reads public governance frameworks from the European Union, NIST in the United States, Singapore, Japan, and Canada. Second, it analyzes provider-side policy statements and scaling-governance materials, with Anthropic used as a salient current articulation of frontier-provider sovereignty. Third, it uses implementation-visible runtime materials to capture the operational fact of heterogeneous execution layers, including managed agent services, observability products, and programmable guardrail systems. Fourth, it incorporates directional market and practitioner evidence to test whether institutional pressure is moving toward upstream centralization or toward downstream authority closure.

The selection logic is deliberately comparative rather than exhaustive. The chosen jurisdictions are not the only ones producing relevant AI guidance, but they are enough to test whether contemporary public governance is converging toward exclusive frontier-provider authority. Anthropic is not the only frontier lab, but it is among the clearest publicly documented cases of a strong upstream gatekeeping position. The runtime materials are not a complete market survey, but they are sufficient to show that enterprise AI action now traverses multiple control surfaces rather than one stable stack.

The comparison proceeds by evaluating rival sovereignty models against six criteria: consequence alignment, context sensitivity, runtime portability, evidentiary closure, frontier-risk visibility, and anti-concentration. These criteria do not ask which actor is most intelligent or most safety-minded in the abstract. They ask which actor is best positioned to exercise legitimate and durable authority over governed enterprise action under real deployment conditions.

This method has a clear limitation. It is stronger at identifying institutional fit than at forecasting long-run equilibrium. Even so, that is the right level of analysis for the present claim. The paper does not need to prove that every future governance regime will become PCAA-like. It only needs to show that under current enterprise conditions, strong frontier-provider sovereignty is less well aligned with the practical requirements of action governance than an action-centered deployer model.

\section{Related Work and the Analytical Gap}

Recent literature already points beyond narrow model-level governance, but it does not yet settle the sovereignty question that concerns this paper.

Goemans et al.\ argue that frontier AI governance cannot rely only on a model-level paradigm because capability progress is increasingly shaped by non-model gains, including inference-time scaling, system-level scaffolds, and asset integration \citep{goemans2026nongains}. Their contribution is important because it weakens the default assumption that upstream model governance alone can remain a sufficient control point as capabilities diffuse into systems and workflows.

Dux et al.\ make a related argument from the enterprise side. Studying organizational deployment of agentic systems, they argue that governance is implemented through concrete architectural and working arrangements that determine what systems are allowed to do, which tools and data they can use, how memory is handled, and how iterative improvement is introduced \citep{dux2026governancebydesign}. This helps move the conversation from abstract AI ethics to operational governance-by-design.

The prior PCAA paper enters this conversation from a more implementation-grounded systems angle. There, \emph{Proof-Carrying Agent Actions} (PCAA) is introduced as a runtime-neutral governance model centered on an action certificate rather than on a vendor-native session record \citep{wang2026pcaa}. The model organizes control around five checkpoints---pre-action admissibility, action open, assumption capture, approval, and outcome closure---and binds those checkpoints to a portable action envelope, approval and runtime receipts, and replay-ready proof \citep{wang2026pcaa}. In other words, PCAA is not presented merely as another policy layer. It is presented as a way of deciding what the primary trust-bearing object should be when heterogeneous agent systems execute across mixed control surfaces.

That prior contribution matters here because PCAA is not yet a widely circulated category term. In this paper, we therefore use ``PCAA'' in a specific and limited sense: a prior runtime-governance formulation in which the governed action itself becomes the portable object of authority, review, and proof \citep{wang2026pcaa}. We do not assume the reader already knows the term, and we do not use it as if it were already a generic industry label.

The strategic disagreement in the present paper is therefore narrower than a general literature review. We do not argue that provider safety is useless, or that model evaluation should be abandoned. We argue that once recent literature has already admitted system-level gains, enterprise architecture, and runtime heterogeneity, the next unresolved issue is where \emph{final authority} should sit. That is the gap this paper addresses directly.

\section{Two Competing Sovereignty Models}

\subsection{Frontier-provider sovereignty}

Frontier-provider sovereignty begins from a serious intuition. The most capable model providers often know more than downstream users about evaluation methodology, internal capability thresholds, scaling behavior, training pipelines, and catastrophic-risk indicators. If one is primarily worried about irreversible frontier mistakes, it is rational to look upstream.

That intuition underlies a growing class of proposals centered on provider-side testing, auditing, release gating, transparency duties, compute governance, and coordinated export controls. Dario Amodei's June 2026 essay is the clearest public statement of this position. There he argues that frontier models should be subjected to technical testing and auditing and that release should be blocked or reversed if safety standards are not met \citep{amodei2026policy}. In related writing, he argues for transparency legislation for frontier AI companies and stronger export controls to preserve strategic advantage and safety buffers \citep{amodei2025deepseek}. Anthropic's Responsible Scaling Policy likewise reflects a provider-centered conception of governance. It is proportional and iterative in presentation, but the decisive objects in the policy remain provider capability thresholds, provider risk reports, provider executive review, and provider-side decisions about escalation and deployment \citep{anthropic2026rsp}.

At its best, this model offers several advantages. It places hard questions close to the capability frontier. It creates a vocabulary for catastrophic risk. It may prevent some classes of irresponsible release that downstream users cannot even detect in time. And because it centers the entity that develops frontier capability, it can respond before dangerous functionality diffuses.

But the same model also creates a structural concentration problem. The provider knows the model best, yet does not bear the full downstream context of each governed action. It does not always know the local regulatory environment, the contractual obligation, the counterparty status, the boundary classification of a destination, the internal approval chain, or the commercial blast radius of a specific act. Provider sovereignty therefore risks a specific failure mode: \emph{authority without consequence}. The provider may be best placed to judge model danger in the abstract while remaining badly positioned to judge whether a particular governed action is acceptable in a concrete enterprise context.

\subsection{Action-centered deployer sovereignty}

Action-centered deployer sovereignty starts from a different intuition: in enterprise settings, governance is not exhausted by model danger. It is equally about authority, legitimacy, evidence, and accountability in context.

Under this view, the question is not merely whether a model is globally safe enough to exist. The question is whether a particular action---an external handoff, a production-side change, an identity elevation, a procurement decision, a regulated communication, a data movement, a customer-impacting update---may proceed \emph{here}, \emph{now}, under \emph{this} organization's policy and consequence structure. The actor that bears those consequences is the actor that should hold final action authority.

PCAA is one architecture built around this idea. Its central design choice is to make the governed action itself the trust-bearing object. Rather than letting authority disappear into model-side refusals, runtime-native traces, or detached workflow tickets, PCAA keeps route, review, and proof on one certificate path. The point is not to replace the runtime. The point is to preserve final authority closure in a tenant-readable form across heterogeneous runtimes.

This model has a different set of strengths. It is sensitive to organizational context. It makes approval semantics explicit. It travels better across runtime churn. It is naturally aligned with audit, procurement, and evidence export. Most importantly, it places decision rights with the entity that will actually answer for the decision.

\subsection{The decisive difference}

The decisive difference between the two models is therefore not a contrast between safety and freedom. Both can value safety. The difference is whether the \emph{sovereign governance object} is located upstream in the frontier provider or downstream in the governed action and its consequence-bearer.

That distinction becomes sharper as agentic systems spread. Once AI moves from answer generation to delegated action, the unit of governance can no longer remain the model alone. The governed action becomes the institutionally meaningful object.

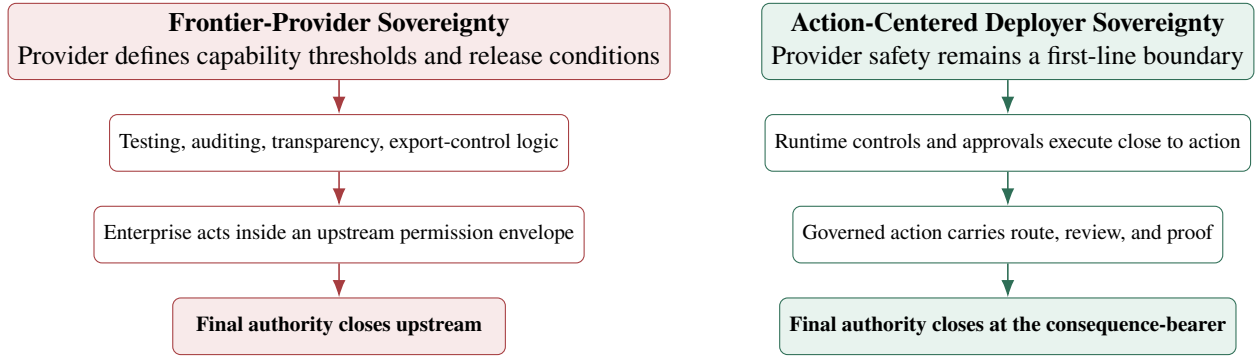
\begin{figure}[t]
\centering
\begin{tikzpicture}[
  node distance=0.45cm and 0.6cm,
  box/.style={draw, rounded corners=3pt, align=center, minimum width=5.1cm, minimum height=0.9cm, font=\small},
  tinybox/.style={draw, rounded corners=3pt, align=center, minimum width=4.4cm, minimum height=0.75cm, font=\scriptsize},
  arrow/.style={-{Latex[length=2.3mm]}, semithick}
]
\node[box, fill=softred, draw=providerred] (p1) {\textbf{Frontier-Provider Sovereignty}\\Provider defines capability thresholds and release conditions};
\node[tinybox, below=of p1, fill=white, draw=providerred] (p2) {Testing, auditing, transparency, export-control logic};
\node[tinybox, below=of p2, fill=white, draw=providerred] (p3) {Enterprise acts inside an upstream permission envelope};
\node[tinybox, below=of p3, fill=softred, draw=providerred] (p4) {\textbf{Final authority closes upstream}};

\node[box, fill=softgreen, draw=deployergreen, right=1.2cm of p1] (d1) {\textbf{Action-Centered Deployer Sovereignty}\\Provider safety remains a first-line boundary};
\node[tinybox, below=of d1, fill=white, draw=deployergreen] (d2) {Runtime controls and approvals execute close to action};
\node[tinybox, below=of d2, fill=white, draw=deployergreen] (d3) {Governed action carries route, review, and proof};
\node[tinybox, below=of d3, fill=softgreen, draw=deployergreen] (d4) {\textbf{Final authority closes at the consequence-bearer}};

\draw[arrow, providerred] (p1) -- (p2);
\draw[arrow, providerred] (p2) -- (p3);
\draw[arrow, providerred] (p3) -- (p4);

\draw[arrow, deployergreen] (d1) -- (d2);
\draw[arrow, deployergreen] (d2) -- (d3);
\draw[arrow, deployergreen] (d3) -- (d4);
\end{tikzpicture}
\caption{The dispute is not whether providers matter, but where final authority closes. Provider-centric models place the sovereign gate upstream; action-centered models preserve provider safety while closing authority on the governed action and its consequence-bearer.}
\label{fig:sovereignty-models}
\end{figure}

\subsection{Evaluation criteria}

To compare the two models fairly, five criteria matter.

\begin{enumerate}
  \item \textbf{Consequence alignment.} The more authority a governance actor holds, the more it should bear the legal, operational, commercial, and reputational consequences of the governed action.
  \item \textbf{Context sensitivity.} A governance model should be able to distinguish between actions that are syntactically similar but institutionally different.
  \item \textbf{Runtime portability.} A governance model should survive heterogeneous runtime estates rather than collapsing into one vendor's session object.
  \item \textbf{Evidentiary closure.} Governance must produce a record that survives later scrutiny: what action was governed, what policy posture applied, what review occurred, and what proof remained.
  \item \textbf{Frontier-risk visibility.} A governance model should respect the fact that providers can often see model-level danger earlier than downstream operators.
  \item \textbf{Anti-concentration.} In a market where frontier capability is increasingly concentrated and transparency is falling, governance theories that rely on upstream concentration deserve unusually hard scrutiny.
\end{enumerate}

\section{A Comparative Reading of Public Frameworks}

\begin{table}[t]
\centering
\scriptsize
\renewcommand{\arraystretch}{1.05}
\setlength{\tabcolsep}{3pt}
\begin{tabularx}{\linewidth}{>{\raggedright\arraybackslash}p{1.45cm} >{\raggedright\arraybackslash}X >{\raggedright\arraybackslash}p{2cm} >{\raggedright\arraybackslash}p{1.65cm}}
\toprule
\textbf{\shortstack[l]{Frame-\\work}} & \textbf{Primary governance emphasis} & \textbf{\shortstack[l]{Downstream\\authority signal}} & \textbf{Sovereignty tendency} \\
\midrule
EU & Strong provider obligations for GPAI models; provider-designed human oversight for high-risk systems & Real deployer obligations remain, but often inside a provider-shaped boundary & Mixed, leaning upstream \\
NIST & Organizational profiles based on user requirements, risk tolerance, and resources & High; deployers and users are explicit governance subjects & Strongly downstream \\
Singapore & Responsible deployment, meaningful human control, and enterprise accountability for agentic AI & Very high; organizations deploying agents remain accountable & Strongly downstream \\
Japan & Trustworthy AI, wide utilization, and lifecycle countermeasures for business users & High; governance is addressed to those who use AI in business contexts & Downstream \\
Canada & Shared code for developers and managers of advanced generative AI systems & Moderate to high; manager guidance operationalizes local oversight and monitoring & Mixed, but distributed \\
\bottomrule
\end{tabularx}
\caption{Public frameworks do not collapse into one model, but the overall direction is toward distributed responsibility rather than permanent frontier-provider sovereignty.}
\label{tab:framework-comparison}
\end{table}
\renewcommand{\arraystretch}{1}

\subsection{European Union}

The EU is the jurisdiction most sympathetic to some parts of the provider-sovereignty instinct, but even here the final picture is mixed rather than absolutist.

On the provider side, the European Commission has issued guidance clarifying obligations for providers of general-purpose AI models, and those obligations entered into application on August 2, 2025 \citep{eu2025gpai}. The architecture is clearly upstream: providers of powerful models carry direct duties around documentation, downstream information sharing, and related compliance infrastructure. Human oversight requirements for high-risk systems also begin from the provider side. High-risk systems must be designed so that natural persons can effectively oversee them during use \citep{eu2026art14}.

Yet the EU does not end with provider sovereignty. Deployer obligations remain real and independent. Article 26 preserves the deployer's freedom to organize its own resources and activities for implementing the human oversight measures indicated by the provider, and deployers carry their own responsibilities around use, oversight assignment, monitoring, and log retention \citep{eu2026art26}. In other words, the EU creates a layered model: providers shape the system-level boundary, but deployers still carry operational responsibility.

The result is not a maximal upstream thesis, but neither is it fully PCAA-like. The EU remains more provider-centric than the other frameworks considered here. Still, even the EU stops short of granting frontier labs unilateral final authority over enterprise action. It distributes obligations across actors.

\subsection{United States: NIST}

The NIST AI Risk Management Framework sits much closer to action-centered deployer sovereignty.

NIST is explicit about its intended audience and posture. The AI RMF is a voluntary resource for organizations designing, developing, deploying, or using AI systems. It is rights-preserving, non-sector specific, and use-case agnostic \citep{nist2023airmf}. Its four functions---Govern, Map, Measure, and Manage---are framed as an organizational risk-management architecture rather than a provider command structure. The generative AI profile extends that logic by emphasizing that profiles should be implemented based on the requirements, risk tolerance, and resources of the framework user \citep{nist2024genai}.

That institutional design matters. NIST does not imagine the frontier provider as the sole sovereign of AI governance. It imagines organizations building governance profiles around their own use cases, risk tolerances, and accountability structures. That is much closer to the PCAA intuition that governance authority should remain legible and owned at the action and organization level.

\subsection{Singapore}

Singapore's 2026 Model AI Governance Framework for Agentic AI aligns especially closely with the action-centered side of the argument developed here.

The framework was launched as a practical guide for enterprises deploying agentic AI responsibly. Its language is direct: humans must remain accountable; meaningful human control and oversight must be integrated into the agentic AI lifecycle; responsibilities must be clearly allocated inside and outside the organization; and governance must include technical controls, processes, and end-user responsibility \citep{imda2026agentic}. In institutional terms, the center of gravity lies with enterprise-operational accountability rather than with model-lab sovereignty.

Singapore's broader AI governance ecosystem reinforces the same logic. The emphasis is on making responsible deployment testable and operational across the actors who actually use and govern AI\@. In sovereignty terms, Singapore is not merely neutral between the two models. It is substantially closer to action-centered deployer governance.

\subsection{Japan}

Japan's emerging AI framework is also substantially closer to deployer sovereignty than to provider sovereignty.

Japan's AI Guidelines for Business are addressed to people who use AI in various businesses. They aim to help them recognize AI risks and voluntarily take necessary countermeasures across the lifecycle \citep{japan2026guidelines}. Japan's 2025 AI Act and 2025--2026 AI Basic Plan reinforce the same orientation. Their emphasis is innovation promotion with risk mitigation, trustworthy AI, and making Japan ``the most AI-friendly country in the world'' \citep{japan2025act, japan2025basicplan}. The institutional posture is not that a frontier provider should rule the field. It is that AI should be widely developed and utilized under a governance model that is agile, trust-oriented, and deployment-conscious.

Japan therefore presents a striking contrast to stronger versions of provider sovereignty. It treats AI governance as a national framework for coordinated use, trust, and practical mitigation rather than as a theory of permanent upstream gatekeeping by frontier labs.

\subsection{Canada}

Canada occupies a middle position, but even there the public trajectory does not support pure provider sovereignty.

Canada's Voluntary Code of Conduct on advanced generative AI systems does begin with developers and managers of advanced systems, which gives it a somewhat more upstream flavor than NIST, Singapore, or Japan \citep{canada2023voluntary}. But the code is explicit that developers and managers have important and complementary roles. The 2025 implementation guide is directed specifically at managers of AI systems and includes best practices for accountability, human oversight and monitoring, transparency, and robustness \citep{canada2025guide}. Canada is therefore operationalizing responsibility downstream as well. It is not fully PCAA-native, but it still rejects the idea that only the frontier provider should hold meaningful authority.

\subsection{Interim judgment}

The comparative pattern is clear. None of the major public frameworks fully instantiates an action-native model like PCAA\@. Most are still organized around systems, lifecycle stages, or actor categories such as provider, deployer, and manager. But the overall trajectory is still unmistakable: public governance increasingly distributes responsibility across downstream operators and consequence-bearers rather than treating the frontier provider as the sole sovereign of AI governance.

This suggests that the policy world is not moving toward strong frontier-lab sovereignty, even when it borrows some upstream controls. It is moving toward layered responsibility. The unresolved step is to make the governed action itself the primary trust object. That is where PCAA contributes something genuinely different.

\section{Enterprise Reality, Market Signals, and Runtime Heterogeneity}

\begin{table}[t]
\centering
\small
\setlength{\tabcolsep}{5pt}
\begin{tabularx}{\linewidth}{>{\raggedright\arraybackslash}p{1.55cm} >{\raggedright\arraybackslash}p{3.2cm} >{\raggedright\arraybackslash}X}
\toprule
\textbf{Source} & \textbf{Observed signal} & \textbf{Why it matters for sovereignty} \\
\midrule
Stanford HAI & Industry produced over 90\% of notable frontier models in 2025; transparency index fell from 58 to 40 & Capability is concentrating while disclosure is weakening, making permanent provider authority harder to justify \\
IBM & Two-thirds of surveyed CIOs and CTOs are accountable for AI systems they do not fully control & Enterprises already bear consequence without matching authority; governance must close locally \\
McKinsey & Persistent gaps remain in strategy, governance, and risk management in the agentic era & The deployment problem is not solved by upstream model safety alone \\
Cisco & Only about 13\% of organizations qualify as AI ``Pacesetters'' & High-performing organizations operationalize governance as a system, not as trust in one provider \\
\bottomrule
\end{tabularx}
\caption{The market evidence points to an accountability problem more than a pure capability problem. That pattern favors action-centered governance.}
\label{tab:market-signals}
\end{table}

The market data strengthens the same conclusion.

Stanford's 2026 AI Index reports that industry produced over 90\% of notable frontier models in 2025, while organizational AI adoption reached 88\% \citep{stanford2026index}. At the same time, the average score on the Foundation Model Transparency Index dropped from 58 to 40 in 2025, with major gaps persisting around training data, compute resources, and post-deployment impact \citep{stanford2026transparency}. Stanford summarizes the moment as a widening gap between what AI can do and how prepared institutions are to govern, evaluate, and understand it \citep{stanford2026gap}.

Those facts cut directly against strong provider sovereignty. If frontier capability is increasingly concentrated in private industry while transparency is declining, then concentrating still more governance authority in providers becomes harder to justify, not easier.

Enterprise surveys point in the same direction. IBM reported on June 8, 2026 that two-thirds of surveyed CIOs and CTOs are being held accountable for AI systems they do not fully control \citep{ibm2026controlgap}. McKinsey's 2026 AI Trust Maturity Survey similarly reports persistent gaps in strategy, governance, and risk management as organizations move into the agentic era \citep{mckinsey2026trust}. Cisco's 2025 AI Readiness Index finds that only about 13\% of organizations qualify as ``Pacesetters,'' and that these higher-performing organizations adopt a disciplined, system-level approach rather than relying on ad hoc experimentation \citep{cisco2025readiness}.

Taken together, these are not signals that enterprises want a frontier lab to rule all consequential action. They are signals of a governance crisis defined by \emph{accountability without authority} and \emph{control without portable evidence}. Enterprises are being held responsible for AI-mediated decisions made across mixed toolchains, mixed runtimes, and mixed approval surfaces. Their actual need is not more abstract frontier rhetoric. It is a governable unit that closes action authority where consequence already sits.

Runtime heterogeneity makes this need sharper. Anthropic's Managed Agents and Microsoft Foundry Agent Service illustrate the rise of provider-managed runtimes. LangSmith illustrates the rise of trace- and observability-centric layers. NeMo Guardrails illustrates the importance of programmable, application-layer control \citep{anthropic2026managed, microsoft2026foundry, langsmith2026observability, nvidia2026guardrails}. Each is useful. None is a universal sovereign. In practice, the enterprise estate will be mixed. Some actions will be intercepted before execution. Others will be reviewed in workflow. Others will only be observed and later reconstructed. Under those conditions, governance must travel above runtime-specific semantics.

\subsection{Practitioner and board-level discourse is moving in the same direction}

The same pattern is now visible in practitioner discourse. Even when the discussion is not framed in academic terms, the center of gravity is moving toward decision rights, ownership, and encoded boundaries rather than toward permanent provider paternalism.

McKinsey puts the issue directly: ``agency'' is not merely a product feature, but a transfer of decision rights. Once agents can plan, call tools, and execute workflows, the central governance question shifts from whether the model is accurate to who is accountable when the system acts \citep{mckinsey2026agents}. The World Economic Forum makes a closely related board-level argument. Boards, it notes, are increasingly reallocating decision rights to autonomous systems while still relying on governance models built for human judgment; the resulting mismatch is not mainly a software problem but a delegation-boundary problem \citep{wef2026boards}.

Operational guidance from large enterprises points the same way. Microsoft describes agent governance as a balance between empowering employees to build and use agents and maintaining control, security, privacy, and compliance through practical governance structures \citep{microsoft2026journey}. IBM argues that autonomy without accountability creates a coordination gap across tools and domains, such that systems can each do what they were designed to do while still producing outcomes misaligned with enterprise intent \citep{ibm2026autonomy}. JetBrains, from the software-tooling side, similarly argues that accountability must be designed into agentic systems through permissions, boundaries, monitoring, and traceability, because enterprises are not only buying AI capability; they are buying trust and operational control \citep{jetbrains2026accountability}.

Public practitioner commentary that circulates heavily on LinkedIn follows the same logic. Andreas Welsch, a LinkedIn Top Voice writing about enterprise agentic adoption, argues that adoption without accountability produces risk, rework, and shadow AI, and that agents must be treated both as ``digital employees'' for governance design and as software for sign-off and control \citep{welsch2026humanedge}. The exact language differs across communities, but the through-line is consistent: as AI systems acquire agency, the market conversation is converging on local accountability and explicit operational authority.

This matters for the sovereignty debate because it weakens a common rhetorical defense of provider-centric governance, namely that the market itself naturally wants stronger upstream gatekeepers. The public evidence is more mixed. What buyers, operators, boards, and enterprise architects repeatedly ask for is not simply safer models in the abstract, but clearer ownership, narrower delegation, stronger boundary-setting, and auditable control where actions actually occur.

\section{Representative Enterprise Scenarios}

The abstract disagreement over sovereignty becomes clearer when mapped onto ordinary enterprise action. The question is not whether a frontier provider can articulate a sensible safety policy. The question is who should decide when a concrete, high-impact action is about to cross a business boundary.

\begin{table}[t]
\centering
\small
\setlength{\tabcolsep}{4.5pt}
\begin{tabularx}{\linewidth}{>{\raggedright\arraybackslash}p{2.35cm} >{\raggedright\arraybackslash}p{2.85cm} >{\raggedright\arraybackslash}p{2.55cm} >{\raggedright\arraybackslash}X}
\toprule
\textbf{Scenario} & \textbf{Decisive local facts} & \textbf{Why provider-side authority is weak} & \textbf{Why action-centered authority is stronger} \\
\midrule
External partner handoff & Destination posture, account type, approved counterparty status, matter sensitivity & The provider cannot reliably know enterprise destination policy or approved transfer context & The deployer owns the transfer boundary, the counterparty relationship, and the approval chain \\
Production-side change & Environment criticality, service dependency, rollback posture, change window, approver role & The provider sees a command or workflow step, not the business blast radius & The deployer knows whether the change is routine, reviewable, or business-critical \\
Identity elevation and exceptions & Role sensitivity, urgency, expiry rules, separation of duties, incident posture & The provider cannot determine who may legitimately grant an exception in local policy terms & The deployer carries the authority model, exception rules, and audit obligations \\
\bottomrule
\end{tabularx}
\caption{Representative enterprise scenarios reveal the same pattern: the decisive governance facts are downstream, local, and consequence-bearing.}
\label{tab:scenario-comparison}
\end{table}

\subsection{External partner handoff}

Consider a document or evidence handoff to an external party: a law firm, forensic vendor, insurer, collections agency, or design partner. The business request often appears mundane and urgent, yet the action becomes sensitive because the agent may choose the transport, the destination, or the account context on its own.

A provider-centric governance model can refuse obviously unsafe patterns, but it usually cannot answer the most important enterprise questions. Is the destination public or restricted? Is the account corporate or personal? Is the counterparty already approved? Was the transfer pre-approved for this matter, or merely tolerated in the abstract? These are local governance facts. The organization that knows and bears them is the deployer, not the frontier provider.

\subsection{Production-side change}

Now consider a production-side change initiated by an agent: rotating a secret, updating a configuration, modifying deployment routing, or changing a payment-related service dependency. To a provider or runtime, this may appear as a shell command, API call, or workflow event. To the enterprise, it may be a revenue-affecting or compliance-relevant act.

Again the core issue is not whether the model was generally cautious. It is whether the organization's approval semantics were honored. Did this action require an explicit reviewer? Did the runtime actually hold before side effects, or was review advisory only? What evidence remains after the change? Provider sovereignty cannot close those questions well because they depend on internal authority and operational consequence, not on generic model capability.

\subsection{Identity elevation and exception handling}

Temporary privilege elevation is a third recurring case. During incidents or business escalations, agents may be asked to grant access, lift restrictions, or create exceptions for contractors and internal users. These are rarely dangerous because of syntax alone. They are dangerous because urgency, ambiguity, and social pressure combine to normalize exceptions that are costly to reverse.

Here the weakness of provider sovereignty is particularly clear. A frontier lab can recommend caution, but it cannot know who in the organization is authorized to grant the exception, whether the target identity is governed by a special policy regime, whether the exception should expire automatically, or what evidence must remain for audit. Those are precisely the conditions under which final authority should remain local and action-centered.

\subsection{Why the scenarios matter}

These scenarios are deliberately ordinary. None depends on a dramatic jailbreak or speculative catastrophic harm. That contrast is central to the argument. The decisive governance failures in enterprise AI are often not spectacular. They are routine actions crossing boundaries under insufficiently clear authority.

If the governed object is the provider session, the governance story fragments across tools, dashboards, and runtime-specific traces. If the governed object is the action, then route, review, and proof can stay attached to the same authority record. That is why these scenarios favor an action-centered model and why they make the sovereignty question operational rather than philosophical.

\section{PCAA as a Governance-Layer Primitive}

The prior PCAA formulation is best read not as a rejection of provider safety, but as a rejection of provider sovereignty \citep{wang2026pcaa}.

\begin{figure}[t]
\centering
\resizebox{0.96\linewidth}{!}{%
\begin{tikzpicture}[
  layer/.style={draw, rounded corners=3pt, minimum width=10.4cm, minimum height=0.95cm, align=center, font=\small},
  note/.style={font=\scriptsize, align=center}
]
\node[layer, fill=softred, draw=providerred] (l1) at (0,0) {\textbf{Provider Safety Layer}: model evaluations, release thresholds, provider-native safeguards};
\node[layer, fill=softblue, draw=paperblue] (l2) at (0,-1.35) {\textbf{Runtime / Execution Layer}: tool controls, traces, holds, workflow enforcement, adapter-specific receipts};
\node[layer, fill=softgreen, draw=deployergreen] (l3) at (0,-2.7) {\textbf{Governed Action Layer (PCAA)}: route, review, approval semantics, replayable proof, tenant-visible authority closure};
\node[layer, fill=white, draw=papergray] (l4) at (0,-4.05) {\textbf{Enterprise Consequence Layer}: audit, procurement, legal accountability, customer assurance, exported evidence};

\draw[-{Latex[length=2.3mm]}, semithick] (l1.south) -- (l2.north);
\draw[-{Latex[length=2.3mm]}, semithick] (l2.south) -- (l3.north);
\draw[-{Latex[length=2.3mm]}, semithick] (l3.south) -- (l4.north);

\node[note, text width=2.25cm] at (-3.3,-5.1) {\textbf{Upstream strength:}\\earliest model-risk visibility};
\node[note, text width=2.25cm] at (3.3,-5.1) {\textbf{Downstream strength:}\\strongest consequence alignment};
\end{tikzpicture}
}
\caption{Where the governance thesis fits. Provider safety and runtime controls remain necessary, but enterprise authority closes most cleanly at the governed-action layer, where approval semantics and proof remain portable across runtimes.}
\label{fig:control-stack}
\end{figure}
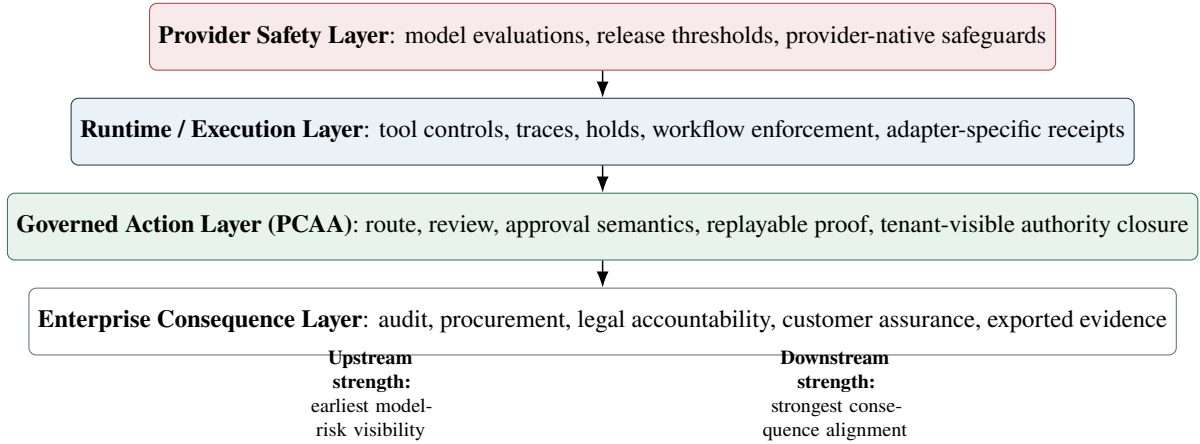

\begin{table}[t]
\centering
\small
\setlength{\tabcolsep}{4.5pt}
\begin{tabularx}{\linewidth}{>{\raggedright\arraybackslash}p{2.15cm} >{\raggedright\arraybackslash}p{2.3cm} >{\raggedright\arraybackslash}p{2.35cm} >{\raggedright\arraybackslash}X}
\toprule
\textbf{Criterion} & \textbf{Frontier-provider sovereignty} & \textbf{Action-centered deployer sovereignty} & \textbf{Interpretation} \\
\midrule
Consequence alignment & Weak to moderate & Strong & Enterprises, not providers, usually bear the direct legal and operational consequences of governed action \\
Context sensitivity & Moderate & Strong & Local policy, destination posture, contractual limits, and approval chains are downstream facts \\
Runtime portability & Weak & Strong & Provider-native control collapses when enterprises run mixed stacks and mixed runtimes \\
Evidentiary closure & Moderate & Strong & Action-centered models attach approval, execution posture, and proof to one authority object \\
Frontier-risk visibility & Strong & Moderate & Providers see model-level capability earliest, which is why provider safety remains necessary \\
Anti-concentration & Weak & Strong & As capability centralizes and transparency falls, governance should not centralize more authority than necessary \\
\bottomrule
\end{tabularx}
\caption{Sovereignty evaluation matrix. The provider-centric model wins on frontier-risk visibility, but the action-centered model dominates on the dimensions that matter most for enterprise governance.}
\label{tab:sovereignty-matrix}
\end{table}

The strongest reading of the provider-sovereignty case is that frontier providers deserve privileged authority because they see dangerous capability earliest. That is true in one domain only: abstract model risk. But enterprise governance is not reducible to abstract model risk. It also includes destination classification, data sensitivity, contractual obligation, customer harm, separation of duties, runtime enforceability, approval semantics, and exportable evidence. These are not model-native facts. They are organizational facts.

A governance primitive built around the governed action therefore has three advantages.

\begin{enumerate}
  \item \textbf{It restores consequence alignment.} The actor deciding whether to proceed is the actor that will bear the aftermath.
  \item \textbf{It restores contextual legitimacy.} A governed action can carry the boundary facts that make the decision meaningful: whether the destination is external, whether the account is corporate or personal, whether approval is advisory or pre-execution, whether the runtime actually enforced the hold, and what evidence survives.
  \item \textbf{It restores portability.} The same authority object can survive runtime churn. This is the difference between a governance model and a feature list.
\end{enumerate}

This is why the route-review-prove logic matters.\@ \emph{Route} gives the enterprise one stable governance vocabulary across changing agent stacks.\@ \emph{Review} ensures that human oversight becomes operational structure rather than a vague slogan.\@ \emph{Prove} closes the action into replayable evidence rather than leaving authority fragmented across model refusals, runtime traces, and disconnected tickets.

Figure~\ref{fig:control-stack} makes the placement claim explicit. The paper does not argue that provider safeguards or runtime-native enforcement should disappear. It argues that these neighboring controls are not, by themselves, the best tenant-visible authority record. The cleanest place for sovereignty to close is the governed-action layer, because that is the first layer where provider caution, runtime posture, approval semantics, and consequence-bearing evidence can be held together without collapsing into one vendor's native object model.

No public framework examined here gives this in full. NIST gives organizational adaptability. Singapore gives operational accountability and meaningful human control. Japan gives deployment-oriented trust and broad utilization. Canada gives complementary developer-manager roles. The EU gives a strong upstream layer with genuine downstream duties. But none of these frameworks yet makes the governed action the sovereign trust object. The PCAA model explicitly attempts to do so by centering authority on the certificate-bearing action rather than on the provider's native session object \citep{wang2026pcaa}.

This is also why PCAA is best interpreted as a governance-layer thesis rather than as a runtime replacement thesis. It belongs above provider-native caution and beside runtime-native enforcement. It is the object that keeps authority closure legible when the underlying runtime can change, the provider can change, and the enforcement depth can change.

\section{Strongest Counterarguments and Boundary Cases}

The argument of this paper becomes more useful if it survives the strongest plausible objections rather than only weaker rhetorical ones. The central counterargument is not that frontier providers are irrelevant. It is that downstream sovereignty may be too fragmented, too inconsistent, and too blind to hidden capability risk to deserve final authority. That challenge should be taken seriously.

\begin{table}[t]
\centering
\small
\setlength{\tabcolsep}{4pt}
\begin{tabularx}{\linewidth}{>{\raggedright\arraybackslash}p{2.55cm} >{\raggedright\arraybackslash}p{3.05cm} >{\raggedright\arraybackslash}X}
\toprule
\textbf{Objection} & \textbf{Strongest version} & \textbf{Implication for the paper's thesis} \\
\midrule
Catastrophic-risk objection & Providers may be the only actors able to detect some dangerous capabilities before release or scaling & Supports strong upstream safety layers, but does not by itself justify provider control over ordinary enterprise action \\
Consistency objection & Local enterprises may apply uneven governance, creating a patchwork weaker than provider-wide rules & Supports minimum upstream floors and shared standards, but not the monopolization of final downstream authority \\
Opacity objection & Enterprises cannot always see latent capabilities, hidden failure modes, or prompt-sensitivity cliffs & Supports better provider disclosure and risk signaling, but enterprise consequence and approval semantics still remain local \\
Convenience objection & Buyers may prefer to outsource governance to providers for speed, cost, or operational simplicity & Tooling can be outsourced; accountability and evidence closure usually cannot \\
Integrated-stack objection & In a vertically integrated environment, provider-native authority may seem complete enough & This is a real boundary case, but the argument weakens as soon as runtimes, obligations, or customer-facing evidence diversify \\
\bottomrule
\end{tabularx}
\caption{The strongest objections mostly support a layered governance settlement rather than exclusive frontier-provider sovereignty.}
\label{tab:counterarguments}
\end{table}

\subsection{The catastrophic-risk objection}

The strongest case for frontier-provider sovereignty is also the fairest one. A frontier provider may see capability jumps, biological or cyber misuse indicators, or scaling behavior that no downstream enterprise can reliably observe in time. If the question is whether a model family should be released, scaled, or granted access to dangerous tool classes at all, strong provider-side authority is not only defensible but often necessary.

That point should be conceded clearly. The present paper does not argue that downstream users should independently overrule frontier safety thresholds whenever convenient. It argues something narrower: the case for provider authority is strongest at the level of \emph{capability access and release discipline}, while the case for deployer authority is strongest at the level of \emph{concrete enterprise action}. Conflating these two levels is what makes provider sovereignty appear stronger than it is. A provider may deserve an emergency veto on catastrophic model deployment without thereby becoming the rightful sovereign of ordinary enterprise approval semantics.

\subsection{The consistency objection}

A second objection is that local governance can be messy. Large enterprises differ in maturity, staffing, risk culture, and implementation quality. From that perspective, centralized provider control appears attractive because it promises uniform rules and fewer weak links.

Uniformity, however, is not the same thing as governance completeness. A globally consistent provider rule still cannot decide whether one bank's outbound transfer, one hospital's disclosure, or one manufacturer's production-side change satisfies its own authority chain, contractual posture, and audit obligations. The most a provider can offer here is a floor: standardized safeguards, minimum control expectations, or provider-native refusal behavior. Those tools are valuable, but they remain incomplete because the decisive local facts are not globally uniform. The correct inference is therefore not that providers should govern everything, but that provider floors should coexist with local authority closure.

\subsection{The opacity objection}

The third objection is epistemic. Frontier providers may know things that downstream operators do not: latent tool-use strategies, instability under long context, hidden policy bypass routes, or dangerous capabilities revealed only under specialized evaluation. If enterprises cannot see these failure modes, why should they hold final authority?

The answer is that hidden capability risk and legitimate action authority are related but not identical questions. Capability opacity strengthens the case for provider disclosure, alerting, eval reporting, and risk-tiered access. It does not erase the downstream fact that the enterprise still decides which counterparty may receive a file, which environment may accept a change, or which exception approver may lift a boundary. In other words, opacity is a reason for \emph{better upstream signal}, not for the permanent transfer of all downstream decision rights.

\subsection{The convenience objection}

Some buyers may genuinely prefer provider sovereignty because it appears cheaper and easier. If the provider offers managed runtimes, built-in safety, and policy defaults, outsourcing governance can feel like operational relief.

But convenience should not be mistaken for resolved authority. Enterprises may outsource infrastructure, monitoring, or policy authoring, yet still remain the actor asked to explain what happened to a regulator, an auditor, a board, or a customer. The harder the post-incident question becomes, the more visible the gap between outsourced controls and non-transferable accountability becomes. For this reason, the relevant distinction is not between ``self-built'' and ``provider-managed'' governance. It is between governance that leaves the consequence-bearer with a tenant-visible authority record and governance that does not.

\subsection{Boundary cases and a layered settlement}

There are real boundary cases where provider-native governance may be close to sufficient. If an organization operates inside one vertically integrated stack, exposes only low-risk internal actions, accepts the provider's native evidence model, and does not need portable approval semantics across runtimes, then provider-centric governance may be serviceable for a period of time.

Yet these cases are less stable than they first appear. Enterprises accumulate second vendors, bespoke tools, external handoffs, regulator questions, and customer-assurance demands. The moment that happens, the object of governance can no longer remain a single provider session without loss of clarity. What looked like a complete sovereignty model is revealed to be a temporary local simplification.

\begin{table}[t]
\centering
\small
\setlength{\tabcolsep}{4pt}
\begin{tabularx}{\linewidth}{>{\raggedright\arraybackslash}p{3.1cm} >{\raggedright\arraybackslash}p{2.45cm} >{\raggedright\arraybackslash}X}
\toprule
\textbf{Governance question} & \textbf{Default sovereign} & \textbf{Reason} \\
\midrule
Should this frontier capability be released, scaled, or granted dangerous tool access at all? & Provider / upstream authority & The provider sees capability thresholds and catastrophic-risk evidence earliest \\
Should this enterprise action proceed under this policy, destination, approval chain, and runtime posture? & Deployer / governed-action authority & The deployer bears the local consequence and owns the operative boundary facts \\
Should execution proceed when both frontier-risk and enterprise-specific consequence are material? & Layered dual control & The correct design is not single-sovereign absolutism but explicit closure at two different levels \\
\bottomrule
\end{tabularx}
\caption{A decision-rules view of the paper's thesis. Different governance questions properly close at different levels.}
\label{tab:decision-rules}
\end{table}

The strongest surviving conclusion is therefore layered rather than absolutist. Provider authority is strongest where the issue is model danger in the abstract; deployer authority is strongest where the issue is whether a concrete action may occur in a real institutional setting. The paper's claim is that current public debate too often lets the first truth swallow the second.

\section{Scope Conditions, Falsifiers, and Threats to Validity}

\subsection{The claim is layered, not absolutist}

The provider-sovereignty thesis should not be dismissed as irrational. In domains involving frontier catastrophic risk, providers may indeed be the first actors capable of measuring some failure modes. A world with no provider-side safeguards, no release discipline, and no frontier testing would be irresponsible.

The argument of this paper is therefore layered rather than absolutist.\@ \emph{Provider safety should remain a first-line boundary.} Frontier providers should continue to invest in evaluation, transparency, incident disclosure, and strong safety practices. Public policy may reasonably impose upstream obligations on them, especially where foundation-model capability creates systemic risk.

But first-line boundary is not the same thing as final sovereign. The moment the question becomes whether a concrete enterprise action may proceed, provider sovereignty becomes structurally misaligned. It is too far from consequence, too far from context, and too fragile under runtime heterogeneity. In that domain, final authority should move downstream.

\subsection{Scope conditions}

This paper also has a deliberate scope boundary. It does not claim that every enterprise should independently decide access to every dangerous model capability. Nor does it claim that PCAA solves catastrophic model governance by itself. The narrower claim is that for enterprise AI action governance, the final authority model should be deployer-centered and action-centered, with provider safety layered underneath rather than enthroned above.

The thesis is strongest under four scope conditions:

\begin{enumerate}
  \item the enterprise operates across mixed runtimes, providers, or control surfaces rather than inside one vertically integrated stack;
  \item the governed action carries local legal, operational, or contractual consequences that the provider does not fully bear;
  \item approvals, exceptions, and exported evidence matter for audit, procurement, or customer assurance;
  \item the organization needs a tenant-visible authority record that survives runtime change.
\end{enumerate}

It is correspondingly weaker in environments where a single provider controls nearly the entire stack, where actions remain low-risk and non-external, or where no meaningful approval or proof obligations exist.

\subsection{Falsifiable implications}

To avoid presenting the argument as pure rhetoric, Table~\ref{tab:falsifiers} states the paper's main empirical implications in falsifiable form.

\begin{table}[t]
\centering
\small
\setlength{\tabcolsep}{4.5pt}
\begin{tabularx}{\linewidth}{>{\raggedright\arraybackslash}p{3.2cm} >{\raggedright\arraybackslash}X >{\raggedright\arraybackslash}X}
\toprule
\textbf{Prediction} & \textbf{What would support it} & \textbf{What would seriously weaken it} \\
\midrule
Runtime heterogeneity increases demand for action-centered governance & Enterprises running mixed stacks ask for portable approvals, tenant-visible authority, and replayable evidence across runtimes & A single provider-native object proves sufficient for approvals, evidence, and audit across mixed enterprise runtime estates \\
Provider sovereignty fits catastrophic-risk governance better than enterprise action governance & Upstream safety rules remain valuable, but downstream organizations still need their own approval and evidence layers & Providers consistently supply context-complete, tenant-legible, locally authoritative governance artifacts for downstream action decisions \\
Public frameworks will continue to distribute responsibility downstream & New guidance continues to emphasize deployers, managers, operators, and meaningful human control & Major public frameworks converge toward exclusive or near-exclusive frontier-provider decision authority \\
Consequence-bearing organizations will resist accountability without authority & Buyers, operators, and boards demand local review, exception handling, and proof closure & Enterprises accept durable liability for AI actions while preferring to leave final authority permanently with frontier providers \\
\bottomrule
\end{tabularx}
\caption{Falsifiable implications of the paper's core thesis. The argument should be weakened if the counterevidence in the rightmost column begins to dominate future practice.}
\label{tab:falsifiers}
\end{table}

\subsection{Threats to validity}

Three threats to validity deserve explicit acknowledgment.

First, the paper relies heavily on public policy documents, provider statements, and implementation-visible materials. These sources are appropriate for analyzing declared governance structure, but they do not fully reveal internal decision processes, unpublished enterprise controls, or private customer behavior.

Second, the paper is closer to the PCAA implementation than to most alternative architectures. That proximity is a strength when discussing practical runtime governance design, but it also creates a risk of interpretive bias in favor of action-centered models. The paper therefore should be read as implementation-informed comparative analysis rather than as a detached market census.

Third, the market evidence cited here is directional rather than dispositive. Survey data, board-level commentary, and practitioner discourse can show where institutional pressure is moving, but they do not by themselves establish that one governance model will dominate all enterprise contexts.

These limits narrow the strongest responsible reading of the contribution. The paper does not prove that every future AI market will converge on PCAA-like governance. It argues that under contemporary enterprise conditions---mixed runtimes, rising accountability pressure, and declining tolerance for opaque authority---action-centered sovereignty is the more robust governance primitive.

\section{Discussion and Implications}

The comparative claim developed in this paper has implications beyond the immediate dispute between provider-centric and action-centric governance models. If the argument is correct, then the next phase of AI governance should focus less on choosing a single sovereign and more on specifying which kinds of authority should close at which layer.

\begin{table}[t]
\centering
\small
\setlength{\tabcolsep}{4pt}
\begin{tabularx}{\linewidth}{>{\raggedright\arraybackslash}p{2.25cm} >{\raggedright\arraybackslash}p{3.4cm} >{\raggedright\arraybackslash}X}
\toprule
\textbf{Audience} & \textbf{Main implication} & \textbf{Operational consequence} \\
\midrule
Policymakers & Separate capability governance from action governance & Provider obligations should remain strong, but should not be drafted as substitutes for deployer approval and evidence duties \\
Enterprise architects & Make the governed action, not the vendor session, the durable trust object & Approval semantics, holds, exceptions, and exported proof should remain legible across runtime churn \\
Platform providers & Compete on signal quality, controls, and evidence interoperability rather than implied sovereignty & Better disclosure, richer receipts, and portable control hooks become strategic differentiators \\
Researchers & Evaluate governance systems by authority closure, not only by model refusal behavior & Comparative work should test which artifacts remain authoritative across mixed stacks and real enterprise workflows \\
\bottomrule
\end{tabularx}
\caption{If the paper's thesis is right, the practical next step is layered governance with clearer authority closure by audience and responsibility type.}
\label{tab:implications}
\end{table}

\subsection{Implications for public policy}

The policy implication is not that provider obligations should be weakened. On the contrary, frontier providers are likely to remain indispensable sources of capability evaluation, abuse signal, incident reporting, and pre-release discipline. The implication is that public policy should avoid treating these upstream duties as complete substitutes for downstream authority.

This distinction matters because many regulatory texts already move toward layered allocation even when they do not use sovereignty language. Providers are asked to disclose, test, document, and gate. Deployers, managers, and operators are asked to organize oversight, assign responsibility, monitor use, and retain evidence. The next step for policy design is to make that layering more explicit. In practical terms, that means regulations and guidance should distinguish at least three questions: who may release or scale a dangerous capability, who may authorize a concrete enterprise action, and what proof must survive after the action occurs.

\subsection{Implications for enterprise architecture}

For enterprise architecture, the paper implies that governance cannot remain a diffuse mixture of model refusals, runtime logs, workflow tickets, and retrospective screenshots. Those artifacts may all matter, but they do not automatically form a coherent authority record. The more heterogeneous the runtime estate becomes, the more costly that fragmentation becomes.

An action-centered design pushes in a different direction. The key architectural requirement is to preserve one tenant-legible object that carries route, review, execution posture, and proof across changing runtimes. This does not require abandoning provider controls or runtime-native safeguards. It requires placing them inside a larger evidence model in which the enterprise can still answer the decisive questions after the fact: what action was proposed, what boundary facts governed it, who approved it, what actually executed, and what evidence remains exportable.

\subsection{Implications for market structure}

There is also a market-structure implication. If governance authority is allowed to collapse into the provider's native session object, then governance concentration may deepen alongside capability concentration. That would make governance quality increasingly dependent on what one provider chooses to expose, preserve, and permit. By contrast, if final action authority remains portable and tenant-visible, then providers still compete intensely, but they compete within a governance environment that is less likely to harden into one firm's preferred control ontology.

This does not eliminate concentration risk by itself. However, it changes where differentiation occurs. Providers can still differentiate on model quality, safety rigor, incident responsiveness, tool ecosystems, and managed runtime quality. What they cannot assume by default is that excellence at the capability layer automatically entitles them to permanent authority over every downstream institutional decision.

\subsection{Implications for future research}

The paper also points to a concrete research agenda. Future work should compare governance designs not only by refusal rates or policy adherence in isolated prompts, but by whether authority remains legible under multi-step execution, mixed-provider estates, delegated approvals, and post-incident review. That research could be empirical, architectural, or organizational. What matters is that it tests governance where enterprises actually experience it: across layered systems, not only at the model boundary.

PCAA is one candidate design in this space, not the only one. Its importance in this paper is illustrative as much as prescriptive. It shows what it looks like to treat the governed action as the primary trust-bearing object. Whether the market ultimately standardizes around PCAA itself, around adjacent certificate-bearing models, or around another portable authority construct remains an open question. What should be less open after this analysis is the need for some form of portable authority closure above the provider-native session.

\section{Conclusion}

This paper has argued that AI governance debates often conflate two distinct claims: that frontier providers may know model behavior best, and that they should therefore hold final authority over consequential downstream action. The first claim may often be true. The second does not follow automatically, especially in enterprise settings where approval semantics, legal exposure, and operational consequences are borne elsewhere.

Recent policy writing associated with Dario Amodei and Anthropic articulates a serious and internally coherent theory of frontier-provider governance. That theory is most persuasive as an argument for upstream safety responsibility under frontier-risk conditions. It is less persuasive as a general answer to the enterprise sovereignty question. When evaluated against public governance trends, enterprise accountability structures, market evidence, and runtime heterogeneity, the provider-centric model appears structurally weaker on consequence alignment, context sensitivity, and portable evidence.

The comparative result advanced here is therefore modest but important: under contemporary enterprise conditions, the more robust governance thesis is that final authority over high-impact AI action should ordinarily sit with the deployer, operator, and consequence-bearer, and that this authority should be made legible through governed actions, explicit review semantics, and replayable proof. In that sense, the PCAA formulation should be read not merely as a product architecture, but as one concrete answer to the sovereignty problem emerging at the center of enterprise AI governance \citep{wang2026pcaa}.

The most defensible settlement is not a simple inversion in which enterprises replace providers as unilateral rulers. It is a layered settlement in which upstream actors retain serious responsibilities for frontier-risk visibility, release discipline, and capability gating, while downstream actors retain final authority over the approval, denial, escalation, and evidentiary closure of concrete enterprise action. Once that distinction is made explicit, the strategic disagreement becomes clearer: the real issue is not whether frontier labs matter, but whether their legitimate upstream role should expand into a generalized claim of sovereignty over the institutions that actually bear downstream consequence.

\end{document}